\documentclass[sigconf, authorversion]{acmart}

\AtBeginDocument{%
  \providecommand\BibTeX{{%
    \normalfont B\kern-0.5em{\scshape i\kern-0.25em b}\kern-0.8em\TeX}}}

\copyrightyear{2024}
\acmYear{2024}
\setcopyright{rightsretained}
\acmConference[CHI EA '24]{Extended Abstracts of the CHI Conference on Human Factors in Computing Systems}{May 11--16, 2024}{Honolulu, HI, USA}
\acmBooktitle{Extended Abstracts of the CHI Conference on Human Factors in Computing Systems (CHI EA '24), May 11--16, 2024, Honolulu, HI, USA}
\acmDOI{10.1145/3613905.3650730}
\acmISBN{979-8-4007-0331-7/24/05}

\begin{document}


\title[A Virtual Environment for Collaborative Inspection in Additive Manufacturing]{A Virtual Environment for Collaborative Inspection in Additive Manufacturing}


\author{Vuthea Chheang}
\orcid{0000-0001-5999-4968}
\affiliation{
  \institution{Lawrence Livermore National Laboratory, Livermore, CA, USA}
  \country{}
  \postcode{94550}  
}
\email{chheang1@llnl.gov}

\author{Brian Thomas Weston}
\orcid{0009-0003-0554-2781}
\affiliation{
  \institution{Lawrence Livermore National Laboratory, Livermore, CA, USA}
  \country{}
  \postcode{94550}
}
\email{weston8@llnl.gov}

\author{Robert William Cerda}
\orcid{0009-0005-2696-6713}
\affiliation{
  \institution{Lawrence Livermore National Laboratory, Livermore, CA, USA}
  \country{}
  \postcode{94550}
}
\email{cerda3@llnl.gov}

\author{Brian Au}
\orcid{0009-0003-3715-3676}
\affiliation{
  \institution{Lawrence Livermore National Laboratory, Livermore, CA, USA}
  \country{}
  \postcode{94550}
}
\email{au7@llnl.gov}

\author{Brian Giera}
\orcid{0000-0001-6543-7498}
\affiliation{
  \institution{Lawrence Livermore National Laboratory, Livermore, CA, USA}
  \country{}
  \postcode{94550}
}
\email{giera1@llnl.gov}

\author{Peer-Timo Bremer}
\orcid{0000-0003-4107-3831}
\affiliation{
  \institution{Lawrence Livermore National Laboratory, Livermore, CA, USA}
  \country{}
  \postcode{94550}
}
\email{bremer5@llnl.gov}

\author{Haichao Miao}
\orcid{0000-0001-6580-2918}
\affiliation{
  \institution{Lawrence Livermore National Laboratory, Livermore, CA, USA}
  \country{}
  \postcode{94550}
}
\email{miao1@llnl.gov}

\renewcommand{\shortauthors}{Vuthea Chheang et al.}

\begin{abstract}

Additive manufacturing (AM) techniques have been used to enhance the design and fabrication of complex components for various applications in the medical, aerospace, energy, and consumer products industries. 
A defining feature for many AM parts is the complex internal geometry enabled by the printing process. However, inspecting these internal structures requires volumetric imaging, i.e., X-ray CT, leading to the well-known challenge of visualizing complex 3D geometries using 2D desktop interfaces. Furthermore, existing tools are limited to single-user systems making it difficult to jointly discuss or share findings with a larger team, i.e., the designers, manufacturing experts, and evaluation team.
In this work, we present a collaborative virtual reality (VR) for the exploration and inspection of AM parts. Geographically separated experts can virtually inspect and jointly discuss data. It also supports VR and non-VR users, who can be spectators in the VR environment.    
Various features for data exploration and inspection are developed and enhanced via real-time synchronization. 
We followed usability and interface verification guidelines using Nielsen's heuristics approach.  
Furthermore, we conducted exploratory and semi-structured interviews with domain experts to collect qualitative feedback. 
Results reveal potential benefits, applicability, and current limitations.      
The proposed collaborative VR environment provides a new basis and opens new research directions for virtual inspection and team collaboration in AM settings. 
  
\end{abstract}



\begin{teaserfigure}
  \includegraphics[width=\textwidth]{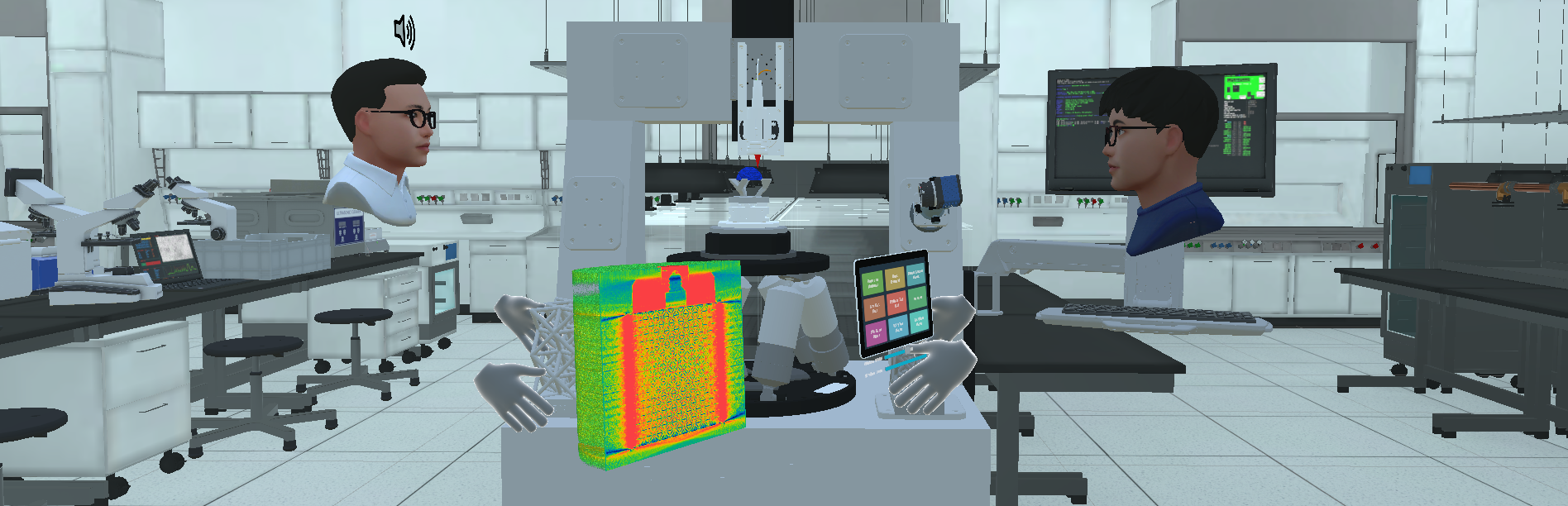}
  \caption{Collaborative users virtually explore and inspect an X-ray CT scan of a complex additively manufactured object. Various immersive visualization and interaction methods for data exploration and inspection of these digital twins were developed and integrated via real-time synchronization.
  }
  \label{fig:teaser}
  \Description{Collaborative users virtually explore and inspect an X-ray CT scan of a complex additively manufactured object. Various immersive visualization and interaction methods for data exploration and inspection of these digital twins were developed and integrated via real-time synchronization.}
\end{teaserfigure}


\maketitle

\section{Introduction}

Additive manufacturing (AM) or 3D printing techniques have been used to enhance manufacturing processes that can fabricate optimal part designs irrespective of part complexity. AM has impacted a wide range of applications in many industrial sectors. 
Compared to traditional techniques, e.g., subtractive manufacturing, in which manufactured parts are created by cutting solid blocks of material, AM provides unique benefits that could offer innovative designs. For example, allowing internal density gradients or stiff, but lightweight infills using complex lattice structures \citep{mazur2016deformation}. 

AM reduces the cost of the prototyping stage and material waste generation compared to conventional approaches. Furthermore, intricate truss structures and similarly complicated geometries produced by AM techniques are not possible via traditional manufacturing \citep{maconachie2019slm}.       
The workflow for AM typically converts computer-aided design (CAD) models to print instructions that are used to fabricate parts, followed by intensive inspection methods of the parts to ensure they meet user requirements. Despite the advantages of AM (and consistent with all novel manufacturing methods), AM systems are nascent and could produce unexpected defects in fabricated parts.
Defects of AM parts are often related to excess or insufficient material, but can also include unwanted material impurities. In the context of defects of lattice structures, trusses can be missing/broken, have an overly rough surface finish, or have unwanted pores within. Therefore, identifying and classifying those defects is of utmost importance.

The inspection of manufactured parts can be categorized into four elements, such as internal defects, external defects, dimensional accuracy, and surface roughness \citep{hassen2018additive}. 
Most inspection analyses rely on \textit{desktop-based systems} with high-resolution computed tomography (CT) scans or other related modalities \citep{klacansky2022virtual, forien2023detecting}. 
Volume rendering and model representations could serve as an approach to explore and analyze the parts. However, desktop-based systems offer limited visualization and interaction opportunities, in particular depth perception, compared to a virtual reality (VR) system \cite{chheang2021collaborative}. 
VR offers great potential and several benefits for AM. Apart from the immersive experience and intuitive interactions, it can be used to enhance the overall workflow. This includes visualization of 3D modeling to identify flaws in the early process, improving the manufacturing process and performance with digital twins by combining virtual-physical representations and data-driven models, and virtual inspection of manufactured parts \cite{attaran2023digital, klacansky2022virtual}.  
In addition, collaboration throughout the design and production process, which often involves separate teams, is crucial. The current process is often realized with in-person meetings and video conferencing, e.g., \textit{Webex} and \textit{Zoom} meetings with screen sharing and PowerPoint slide presentations. 
Collaborative VR can be used to improve several aspects of these processes, including team communication, intuitive interactions, and real-time synchronization for collaborative tasks over distance.    
However, current integrations for AM parts exploration and inspection are still underutilized and are not flexible concerning the aim of the process, e.g., collaborative scenarios and team setup \citep{pirker2022immersive, oppermann2023industrial}. 

In this work, we present a VR environment that allows multiple users to collaborate, explore, and inspect AM parts (see \autoref{fig:teaser}). 
The environment allows users to join either in a co-located or remote space.
It can be used to improve AM processes ranging from data exploration and visualization to in-depth team discussion. Various features are developed and enhanced in a real-time synchronization manner. 
We followed Nielsen's heuristics approach \cite{nielsen1994enhancing, nielsen1994usability} for usability and interface verification. 
Furthermore, we conducted exploratory and semi-structured interviews with six domain experts.
Qualitative feedback regarding the potential benefits, applicability, and current limitations are collected and described. 
The proposed VR environment opens new directions for team communication and collaboration in AM.  
Our contributions are the following:
\begin{itemize}
    \item Design and implementation of a collaborative VR environment to enhance team communication and collaboration on inspecting AM parts.
    \item Adaptation and enhancement of data visualization and synchronization in the collaborative VR environment.
    \item Exploratory analysis aimed at determining potential benefits, applicability, limitations, and research directions.
\end{itemize}

\begin{figure*}[t]
    \centering
    \includegraphics[width=\textwidth]{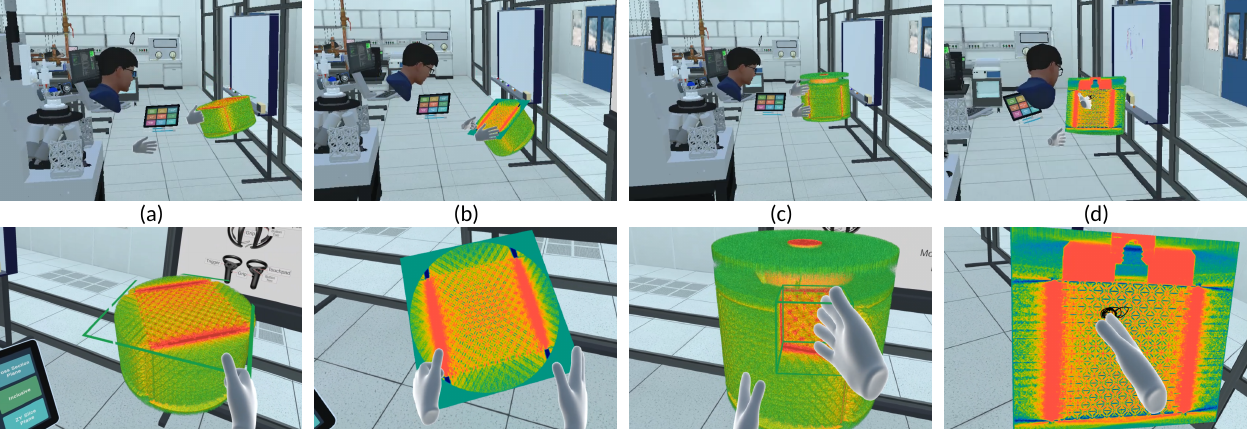}
    \caption{Interaction possibilities with the volumetric data of AM octet lattice structures from the third-person view (top) and first-person view (bottom): (a) the users can use a cross-section plane to explore the dataset from different angles, (b) they can use an axis slicing view to inspect it from a specific axis (axial, coronal, and sagittal), (c) they can explore the region of interest with cutout features, and (d) they can also draw annotations on the dataset.
    }
    \label{fig:inspection}
    \Description{Interaction possibilities with the volumetric data of AM octet lattice structures from the third-person view (top) and first-person view (bottom): (a) the users can use a cross-section plane to explore the dataset from different angles, (b) they can use an axis slicing view to inspect it from a specific axis (axial, coronal, and sagittal), (c) they can explore the region of interest with cutout features, and (d) they can also draw annotations on the dataset.}
\end{figure*}

\section{Related Work}


In recent years, there has been a growing interest in creating virtual analogous physical components, e.g., digital twins, to aid the overall AM process \cite{attaran2023digital, kritzinger2018digital}. Compared to desktop-based systems, VR offers immersive visualization, intuitive interactions, realistic rendering, navigation, and other functionalities. 
\citet{mathur2023designing} proposed a virtual training environment of AM for design students in VR. Their results show that VR can be used in training programs and curricula to improve design and problem-solving skills.

\citet{klacansky2022virtual} presented a virtual inspection tool for comparing the CAD model of designed parts with the CT scan of the corresponding AM parts. They concluded that using immersive 3D rendering is more effective and allows measurements for 3D metrology in AM straightforwardly compared to existing tools. 
\citet{pirker2022immersive} reviewed application scenarios that benefit from a combination of VR and digital twins.
Similarly, \citet{del2023digital} provided a systematic literature review to investigate the intersection of VR and digital twins in industrial contexts.
Their results demonstrate the potential and opportunities of VR. The potential benefits include a low-cost approach, direct and natural interactions, advanced data collection, advanced visualizations, and remote collaboration.  



While VR has the potential to provide better visualization and interactions, most of the systems are designed as a single-user approach. On the other hand, collaborative VR allows multiple users to join and collaborate simultaneously in the same shared virtual environment \cite{chheang2024advanced}. It can be used to support communication, team training, group discussion, and particularly collaborative work between multidisciplinary teams~\cite {chheang2020toward}. 
\citet{stacchio2022empowering} presented a framework by integrating user annotations to support the design in the manufacturing context for both augmented and virtual environments. 
\citet{oppermann2023industrial} introduced a mixed reality tool to support remote maintenance in the collaborative environment. They also discussed some of the practical challenges and complexities during the development of collaborative interactions, such as remote rendering, mixed reality hardware limitation, and data synchronization. 

\citet{tahemaa2019digital, kuts2020digital} proposed a VR environment for industrial digital-twin robot synchronization. They concluded that the concept of a digital twin in a collaborative VR environment is practically viable, and can be used for industrial applications in the near future.   
\citet{havard2019digital} developed an environment for industrial design and assessment between the digital twin and VR on a human-robot collaborative use case. They proposed a solution using client-server architecture and real-time machine-to-machine communication for data exchange.  
Collaborative data management could play an essential part in digital twin for AM \cite{xiong2023human}. \citet{liu2022digital} presented a framework to support the development of digital-twin data management in different product lifecycle stages. It includes the stages for product design, process planning, manufacturing, post-processing, and quality measurement.

Compared to previous work, the proposed collaborative VR environment offers unique advantages for team-based inspection and collaboration ranging from immersive data visualization and interactions to team discussion in a real-time synchronization manner.


\section{Collaborative VR for AM Inspection}

In the following sections, we describe the design and implementation of the proposed collaborative VR environment, including requirement analysis, system architecture, system features, and interface verification.

\subsection{Requirements Analysis}

We identified the requirements based on the meeting and discussion with AM experts from the advanced manufacturing laboratory (AML). The following requirements were determined to develop and optimize the proposed environment.

\begin{itemize}
    \item[\textbf{R1}] For inspecting complex structures of the AM part, it is essential to involve multiple experts, e.g., experts from the design, production, and imaging team. The VR environment has to support and allow multiple users to join in the same shared virtual environment whether they are in co-located or remote physical space.  
    
    \item[\textbf{R2}] Since X-ray CT scans are crucial for inner structure inspection, the visualization and synchronization techniques for collaborative data exploration and inspection should be adapted and enhanced. This includes volume rendering and mechanisms for real-time synchronization.
    
    \item[\textbf{R3}] Handling and rendering volumetric data in VR could affect the user experience. The system should provide features for inspecting volumetric data, e.g., slicing, cutout, and annotations, while also maintaining the system's performance to avoid discomfort for VR users. 
\end{itemize}

\subsection{Collaborative Exploration and Inspection} 


Once the users join the virtual environment, they can start exploring the environment as well as interaction techniques, e.g., navigation and UI interactions. 
We used the dataset of AM octet lattice structures ($1200\times1200\times1200$)~\cite{miao2022data} as an example for collaborative inspection. The defect in AM can potentially result in a bent strut, broken strut, missing strut, thin structs caused by insufficient material, or dross defect caused by excessive material. 

\autoref{fig:inspection} shows inspection features and interaction possibilities in the virtual environment. 
The users can use the virtual tablet to facilitate exploration and inspection features, such as loading datasets, a cross-section plane, box and sphere cutouts with inclusive and exclusive features, and slicing views with different axes (axial, coronal, and sagittal plane).
The users can use the cross-section plane to slice the volume data from different angles. It can be moved and rotated freely based on the interaction with the VR hand (see \autoref{fig:inspection}a). 
To get the view from the exact axis angle, the users can use the axis slicing view to change the image slices (see \autoref{fig:inspection}b). 
Furthermore, they can use the box or sphere cutout to explore the region of interest of the dataset (see \autoref{fig:inspection}c).
There are options to change the volume intensity, i.e., window width and level, on the virtual tablet as well.
All interactions are synchronized in real time. 
Besides the exploration features, the users can draw lines to point to the region of interest and initiate the discussion (see \autoref{fig:inspection}d). 
A virtual whiteboard was also implemented allowing the users to draw and illustrate their ideas, which could be essential for discussion with their collaborators.

\begin{figure*}[t]
    \centering
    \includegraphics[width=\textwidth]{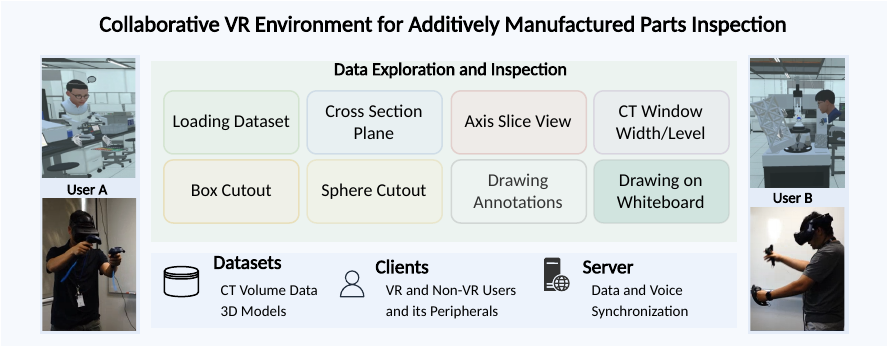}
    \caption{Overview of the collaborative VR environment for AM parts inspection. The users can connect to join in the virtual environment either in a remote or shared physical space. While our aim is focused on VR users, non-VR users can also connect in spectator mode. Interactions for data exploration and inspection are synchronized in real time between users. We use a client-server architecture for data and voice synchronization. }
    \label{fig:systemarchitecture}
    \Description{Overview of the collaborative VR environment for AM parts inspection. The users can connect to join in the virtual environment either in a remote or shared physical space. While our aim is focused on VR users, non-VR users can also connect in spectator mode. Interactions for data exploration and inspection are synchronized in real time between users. We use a client-server architecture for data and voice synchronization.}
\end{figure*}

\subsection{System Architecture}


\autoref{fig:systemarchitecture} shows an overview of the proposed collaborative VR environment. 
We used a \textit{Unity} game engine (Unity Software Inc., CA, USA) for the development environment.
A client-server architecture is used to handle data synchronization. We used \textit{Photon networking -- PUN 2} (Exit Games GmbH, Germany) to provide the load-balancing service and shared sessions between the clients. Additionally, voice communication is realized using \textit{Photon Voice 2}. 
Based on the client's computer capabilities and network conditions, the performance of collaborative mode could be affected. However, major factors, including delay, bandwidth, jitter, and packet loss were considered and optimized during the development. 

We avoid sending a large amount of data over the network. 
This approach stores the object states and performs rendering locally on the client side, while sending updates, e.g., position and rotation, to other clients through the server. We also ensure that it happens only during an interaction, e.g., when a user is grabbing an object. Furthermore, a communication mechanism, \textit{remote procedure calls (RPC)}, was utilized to handle the events, send requests, and distribute the data.

The proposed environment can facilitate different user roles, including VR and non-VR users. While the main focus is on VR users, non-VR users can also enter the VR environment as spectators with the mouse and keyboard. The idea is to allow other users, e.g., a supervisor, to quickly join in the environment to provide feedback or initiate the discussion. 
We used a \textit{Ready Player Me} toolkit to design the user avatars. The avatars include personalized heads, animated hands, name tags, and voice icons, which appear on top of the avatar's head when they engage in voice communication. 
A virtual tablet is used to allow users to interact and explore developed features. The datasets are loaded with a predefined color transfer function at the initial stage. We adapted and enhanced \textit{UnityVolumeRendering} to perform volumetric rendering in \textit{Unity}. 
\textit{VTK} color maps were integrated to enhance the process of data exploration. 
The virtual environment was designed to reflect the advanced manufacturing lab to provide an impression of the surrounding environment, including 3D printers and related devices.

\subsection{Usability and Interface Verification}


Usability and interface verification are two important aspects of user interface (UI) design and development. It is the process of determining whether the interface meets the specified requirements and conforms to the standards. 
One of the most widely used techniques for verifying interface and usability is Nielsen's usability heuristics \cite{nielsen1994enhancing, nielsen1994usability}. 
We followed the guidelines and identified the potential factors to avoid common usability problems and to improve the environment design.
The ten heuristics applied to our scenario are listed in the following.

\begin{enumerate}
    \item[i] \textit{Visibility of system status}: there are a number of developed features ranging from avatar design to interactive elements in the VR environment that keep users informed about what is happening, e.g., through the animated hand models while interacting with the virtual objects and a virtual tablet showing the possible system features. 
    
    \item[ii] \textit{Match between system and the real world}: we aim to mimic the real-world AM environment to a great extent. We designed and integrated models of AM components, including 3D printers and truss structures. Users can navigate and explore the environment by simply walking within the designated area or using the teleportation technique.      
    
    \item[iii] \textit{User control and freedom}: users can freely interact with the objects and hand them over to other users. The system allows users to use data exploration features, i.e., cross-section plane and cutout function. Additionally, the users can draw and delete lines on the data representation. 
    
    \item[iv] \textit{Consistency and standards}: we designed the VR environment in realistic life-size measurements, including models of the 3D printer and user avatars. The users may scale the objects, however, the measurements of the dataset are consistent with the original dataset.   
    
    \item[v] \textit{Error prevention}: we followed the iterative development process, which include a number of tests, and identified major factors, e.g., frame rate in VR, to monitor potential errors. 
    
    \item[vi] \textit{Recognition rather than recall}: as we designed the virtual tablet and integrated an instruction board, it would minimize the user's memory load. Moreover, interaction mappings with the controllers were designed in a way that users can easily understand, e.g., grabbing the virtual object with the grip button. 
    
    \item[vii] \textit{Flexibility and efficiency of use}: the environment was designed for both VR and non-VR users. VR users can use VR headsets and controllers, while non-VR users can join the environment as the spectator mode by using a mouse and keyboard. 
    
    \item[viii] \textit{Aesthetic and minimalist design}: developed features and the interface are simple and straightforward, e.g., users can use the touchpad to navigate and interact with the UI in the same way. Additionally, the users can use the cross-section plane together with the axis-slicing view to explore the data. 
    
    \item[ix] \textit{Help users recognize, diagnose, and recover from errors}: the interactions are synchronized in real time to all users. This could allow users to recognize their actions and collaborate effectively. There are also indications related to actions, e.g., green and red lines during navigation showing users the valid destinations.
    
    \item[x] \textit{Help and documentation}: as we intended to design the system for expert users, they may be familiar with datasets and processes. Some experts were also involved in the development process. However, the support and additional training are needed. 
    The documentation is expected to be delivered at the end of the project.

\end{enumerate}

\subsection{Apparatus} 
We used two high-performance computers for testing. They were the same model \textit{RAZER BLADE 16}, equipped with a \textit{13th Gen Intel Core i9-13950HX} processor with 32 cores, an \textit{NVIDIA GeForce RTX 4090} graphics card with 16GB of \textit{VRAM}, and 32GB of \textit{RAM}.
During the development, \textit{HTC VIVE Pro} VR headset was used.
For evaluation, one user used a \textit{VIVE XR Elite} VR headset with a resolution of $1920\times1920$ pixels per eye (offering a combined resolution of 3840 x 1920 pixels), a 110-degree field of view (FOV), and a refresh rate of 90Hz. 
The other user used a \textit{Meta Quest 3} headset, featuring a resolution of $2064\times2208$ pixels per eye, a 110-degree FOV, and a refresh rate of up to 120Hz.

\section{Expert Feedback}

We evaluated the proposed environment with six domain experts (E1--E6). 
Three of them were computer scientists working with AM and the other three were engineers/material scientists.
One had nine years of working experience, three rated themselves between 3--5, and the other two had between 1--2 years of working experience. Regarding VR experience, one had none, four had little experience (using VR a few times), and one had much experience (using VR several times).
We conducted exploratory and qualitative interviews to obtain their feedback about benefits, applicability, and current limitations as well as research directions. 
After explanation and introduction, the participants were asked to join in the virtual environment with one of the researchers. The participants were guided through all the features and interaction possibilities. Moreover, a think-aloud protocol was used during the session.   
After that, we conducted a semi-structured interview to collect qualitative feedback. During the interview, one researcher (interviewer) took notes of all relevant comments. Those comments were then collated in a database, and redundancies were removed.

\paragraph{Team-based Inspection and Collaboration}

All participants stated that current approaches for their team communication and collaboration are realized with in-person meetings, which require back-and-forth travels, and video conferencing by using screen sharing, screenshots, and even using \textit{Microsoft} paint to visually explain ideas. 
E5 said \textit{``We use WebEx video conferencing with screen sharing and PowerPoint slides to give presentations to our collaborators. Being able to visualize the machine, hardware, and parts and handle it in the virtual context is a game changer.''}
While in-person meetings are crucial in reducing misunderstandings, this is not always the case, especially during the \textit{COVID-19} pandemic.   
The experts confirmed that using collaborative VR is extremely helpful in visualizing and guiding their collaborators through the data, machine, and installation setups, particularly with new experimental results. It offers real-time synchronization, intuitive interactions, and team engagement, which are beneficial for communication and discussion cycles, in particular between design and production agencies.
E2, E3, E5, and E6 clearly stated that it is invaluable during the AM design phase by allowing designers, engineers, and stakeholders to collaboratively explore various design aspects, fit, form, and functions. This leads to better-informed decisions, optimized designs, and reducing the need for physical prototypes and travels.

\paragraph{Exploration and Inspection Features}
Regarding features for data exploration, E4, E5, and E6 commented that scaling and slicing the data in VR are the most useful features for them because they could easily explore and inspect the printing artifacts, especially with their collaborators. 
E3 expressed that drawing annotation was very helpful in highlighting the point of interest for discussion. However, it would be beneficial to allow users to draw with different colors or a unique color for each user. 
Similarly, E6 suggested adding the drawing of annotations on the slicing plane, this could avoid drawing through to the other side in 3D space.
VR hand representations with animations were assessed as supportive while grabbing virtual objects. Nonetheless, adding VR tooltips or a help button could help users easily understand the button-mapped functionalities.
E1 mentioned that visualizing multimodal data, including printing toolpath, sensing data, and captured images during the printing process would be very helpful. By visualizing each step, the team can identify bottlenecks, optimize workflows, and enhance efficiency accordingly.
Moreover, E3 and E4 commented that providing additional metadata information, e.g., strut diameter, and a measurement tool would be advantageous for measuring and understanding the data.

\paragraph{Applicability}
For applicability, E2 mentioned that the adaptation of technology could be easily adapted by their team. E3 also added that there seem no issues with senior technicians since it is just a learning curve, and obviously, the system can be used to enhance their process.   
E1 and E2 suggested investigating the approach to handle data privacy and security for collaboration over geographical locations.
E3 and E6 expressed that incorporating it into their workflow might face challenges due to logistics and accessibility. 
Safety while using VR in a limited space in their workplace should be considered as well.
The experts also confirmed the advantages of employing collaborative VR for team training and skill development in AM processes, such as machine setup, material handling, and post-processing without the potential risk of damaging expensive equipment.

\section{Discussion}

Collaborative VR can enable users to design, test, and optimize AM components in a realistic and immersive way, as well as to collaborate with other users across different locations and disciplinary.
We demonstrated the prototype to the experts following guided exploration and interviews, and the sessions were informal. One researcher joined in the virtual environment with them in each session. 
They confirmed its benefits and usefulness, in particular the collaborative capabilities (\textit{\textbf{R1}}). The environment opens a new approach to exploring and qualifying AM parts. 
In the beginning, they were struggling with the use of controllers and interaction techniques. However, it was a learning curve; after they familiarized themselves with the technology, the performance was faster. They also stated that they learned it quickly because another user was showing the interaction possibilities via real-time synchronization. 
VR is still a nascent technology; thus, allowing users to join as a spectator mode with a 2D desktop was considered as helpful. 

All participants were positive about the proposed VR environment. 
When it comes to the question of integrating into their workflow, they highlighted the potential viability. 
The proposed environment can be used for collaborative data exploration (\textit{\textbf{R2}}). They confirmed that using collaborative VR is of utmost useful to enhance their process, e.g., visualizing and guiding their collaborators through data and machine setup. 
There were no issues regarding discomfort or any motion sickness in the VR environment, which could indicate that the system rendering performance is acceptable (\textit{\textbf{R3}}). However, quantitative assessments are needed in future work.
Apart from developed features, integrating more scenarios that go beyond a research prototype would be interesting, e.g., data streaming from a corresponding server, measurement, and alignment tools. Data streaming could play an essential part in their workflow because it could greatly enhance the process and data security. 
One potential approach is to develop a custom WebAPI and utilize \textit{OpenViSUS} for data management. It also could be beneficial for data storage and sharing in the collaborative VR environment as well. 

Besides the inspection features, investigating natural and intuitive VR interaction techniques, including high-precision drawing and annotations using input devices, e.g., VR pen, could be essential in increasing precision, making the annotation and drawing smoother and engaging, and reducing mental load~\cite{chen2022vrcontour, allgaier2022comparison}. 
For measurement, it would be interesting to integrate a tool measuring the length between two points in a correct unit.  
Furthermore, alignment techniques, e.g., \textit{AMP-IT} and \textit{WISDOM} \cite{rodriguesamp} could be essential to provide a precise manipulation and alignment between objects.

VR is primarily used with visual and auditory senses. 
However, AM involves physical properties, such as material strength, thermal conductivity, elasticity, sensor data, and archived datasets. Representing this information accurately in the virtual environment could remain a challenge.
AM also comprises various techniques, e.g., fused deposition modeling (FDM), selective laser sintering (SLS), direct ink write, laser powder bed fusion, and others, with distinct parameters and processes. Replicating and simulating all these processes in VR could be highly complex and may require simplifications.
There are also other challenges to utilizing collaborative VR for the AM process. 
These include high-performance hardware and network requirements.
Collaborative VR requires high bandwidth and low latency to ensure smooth and synchronous communication among users. 
Additionally, it could pose some human factors issues, such as discomfort, which may affect the user experience and performance while wearing it for a longer time. 
Future work should investigate the effects of collaborative VR on the user's cognitive, affective, and behavioral outcomes in AM with an extensive user evaluation. 

Integrating artificial intelligence (AI), e.g., generative AI assistant~\cite{chheang2024towards}, and cloud-based sensing data center, including \textit{Omiverse} and \textit{AWS Digital Twins} in the virtual environment would be interesting. 
Moreover, exploring the integration of multimodal data and progressive refinement interfaces to enhance the realism and fidelity of AM could be crucial as well.

Compared to 2D interfaces with a variety of systems and features, including video conferencing and web-based collaborative tools, VR is still nascent and evolving. However, it is an emerging technology and is increasingly developed and used in various applications. 
VR offers several unique advantages, such as immersive experience, improved depth perception, intuitive interactions, and enhanced collaboration.
Apart from immersive VR, mixed reality (MR) has also been used and applied to various fields. It provides an alternative approach to visualizing and interacting with virtual objects in the physical world. Investigating in MR, e.g., situated analytics and collaboration, would be interesting for future work to provide and enhance understanding and situational awareness.

The proposed collaborative VR offers a new approach and pushes the boundaries of current inspection methods in AM. It also provides new opportunities to integrate with digital twins and has the potential for other related fields, such as medical aerospace, and consumer products industries.

\section{Conclusion}

We have presented a collaborative VR environment for AM parts exploration and inspection. We developed and enhanced collaborative features ranging from data exploration to team discussion. The proposed environment allows users, particularly design and production agencies, to communicate and enhance their workflow accordingly. 
We followed the guidelines of usability and interface verification using Nielsen's usability heuristics. 
Moreover, qualitative feedback from experts was collected and summarized. 
To sum up, collaborative VR is a promising tool for advancing AM and opening new research directions in related domains.

\begin{acks}

This work was performed under the auspices of the U.S. Department of Energy by Lawrence Livermore National Laboratory under Contract DE-AC52-07NA27344. 
The project has been supported by LLNL LDRD (23-SI-003). The work is reviewed and released under LLNL-PROC-859114. We are thankful to study participants for their support and valuable feedback. 

\end{acks}

\bibliographystyle{ACM-Reference-Format}
\balance
\bibliography{Bibliography}


\end{document}